# Thermal expansion of organic superconductor k-($D_4$-BEDT-TTF)$_2$Cu{N(CN)$_2$}Br. Isotopic effect

A.V. Dolbin[1], M.V. Khlistuck[1], V.B. Eselson[1], V.G. Gavrilko[1], N.A. Vinnikov[1], R.M. Basnukaeva[1], V.A. Konstantinov[1], Y. Nakazawa[2]

[1] *B. Verkin Institute for Low Temperature Physics and Engineering, NAS of Ukraine, 47 Nauky Ave., Kharkov, 61103, Ukraine,* [2] *Department of Chemistry, Graduate School of Science, Osaka University, 1-1 Machikaneyama-cho, Toyonaka, Osaka 560-0043, Japan*

e-mail: dolbin@ilt.kharkov.ua

**Abstract.**

Linear thermal expansion coefficient (LTEC) of single crystal κ-($D_4$-BEDT-TTF)$_2$Cu[N(CN)$_2$]Br was studied across the crystal layers in the temperature range 2-290 K using the method of precise capacitive dilatometry. Below $T_c$ = 11,6 K the LTEC of the sample had a small negative value, which is apparently due to the transition from the paramagnetic metal in the superconducting state. There was a bend of temperature dependence of the LTEC, which shows broad peak around 40 K and can be attributed to the elastic lattice anomaly around the end-point of Mott boundary. A sharp jump in the LTEC values and hysteresis was observed in the area of $T_g$ ~ 75-77 K, what is likely explained by the transition in a glass-like state. The isotope effect in the thermal expansion is discusses, which manifested itself in a shift of the phase transitions in comparison with fully deuterated BEDT-TTF sample.

# 1. Introduction.

Low-dimensional organic molecular conductors and magnetic materials are a new type of conductive compounds, which are characterized by a delicate balance between the various types of electronic instabilities [1-2]. Their study yielded some interesting results in various areas of solid state physics (metal-insulator and metal-superconductor transitions due to electron correlations, nesting and reconstruction of Fermi surface with low-dimensional characters, drastic changes in transport properties induced by a magnetic field, the angular magnetoresistance oscillations through interlayer coupling, etc.) There are currently over 100 low-dimensional synthesized organic superconductors, which are inherently radical cation salts. The maximum temperature of the superconducting transition in them reaches 13 K.

Although the search for new low-dimensional molecular superconductors in class of ion-radical salts is continued, the interest in this area in recent years largely shifted to the creation of multi-functional materials combining the (over) conductivity with other physical properties. Design and synthesis of hybrid molecular systems, that combine two or more physical properties, such as (super) conductivity, magnetism, photochromism, ionic conductivity, nonlinear optical properties, etc., is currently among the most rapidly developing areas in the chemistry and physics of new materials. The combination of these properties in a crystal lattice and their synergism can lead to new physical phenomena and applications in molecular electronics. The unusual properties of molecular conductors and magnetic materials are due to the peculiarities of the crystal structure of these substances. Organic (super) conductors and magnets on the basis of the cation - radical salts are quasi one-dimensional or quasi-two-dimensional systems, whose electronic structure is characterized by the presence of conductive packages or layers of organic (metal organic) $\pi$-electron donors linked by weak electrostatic intermolecular interactions [1-2].

The basis of one of the largest groups of organic conductors is bis (ethylenedithio) tetrathiafulvalene (BEDT-TTF), which was obtained by modifying a TTF in 1978 [3], and is a good electron donor. BEDT-TTF molecule consists of two five-membered and two six-membered rings, each of which contains two sulfur atoms (see Fig. 1a). The same figure also shows the crystal structure of the compound.

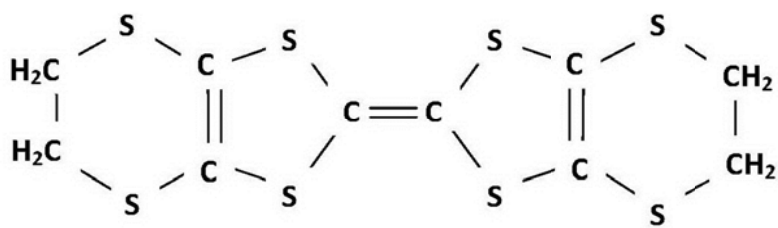

a)

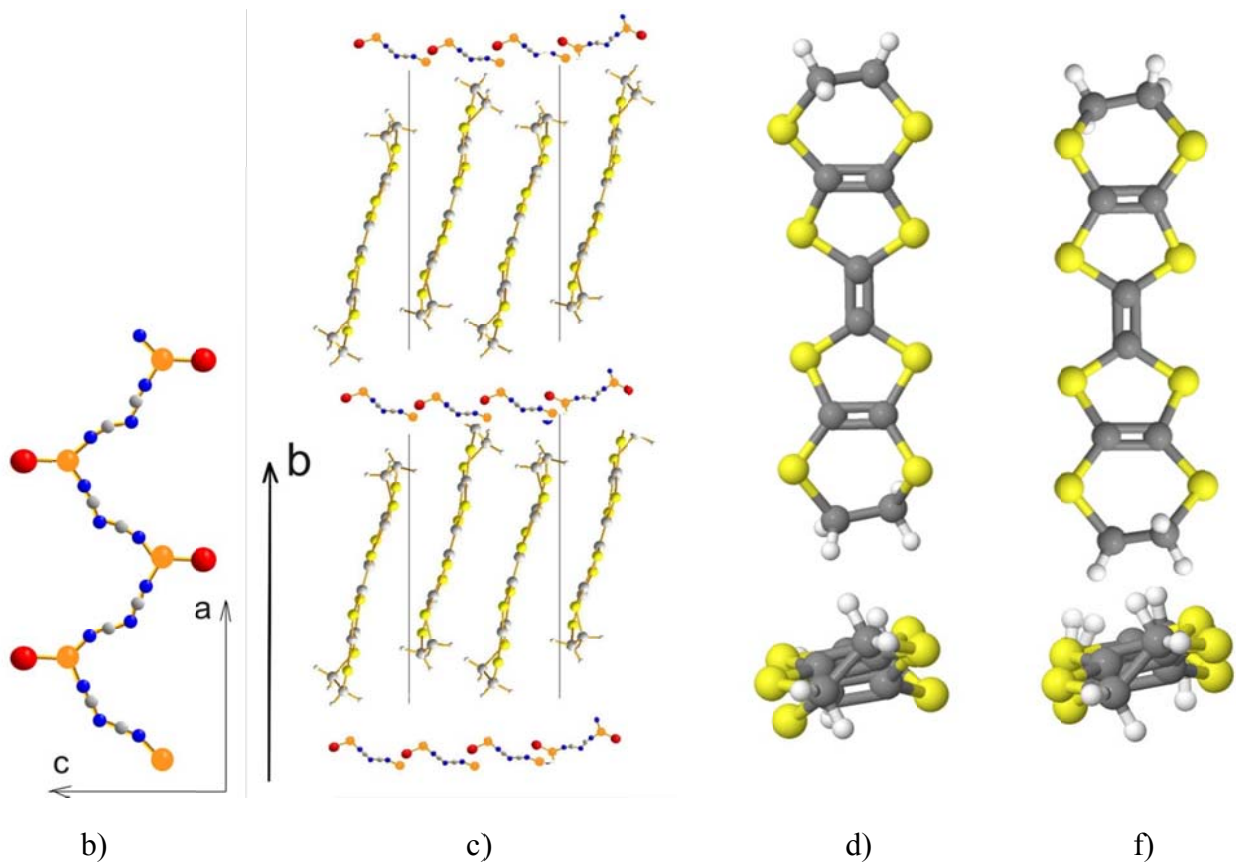

| b) | c) | d) | f) |

Fig. 1. a) BEDT-TTF molecule; b) polymer anions Cu[N(CN)$_2$]Br, arrows indicate the direction in the crystal lattice of κ-(ET)$_2$Cu[N(CN)$_2$]Br; c) crystal structure of κ-(ET)$_2$Cu[N(CN)$_2$]Br, side view showing the sequence of conducting ET layers and insulating anion sheets; d) schematic view of the relative orientations of the ET molecules ethylene end groups (EEGs) in the staggered (d) and eclipsed (f) conformation, respectively [4].

Unpaired π-electrons of carbon and sulfur atoms of BEDT-TTF molecules form a set of π-orbitals, perpendicular to the plane of the molecule [1-2]. Due to such configuration BEDT-TTF molecules can form dense dimers, stacks and other aggregates in crystal in which overlapping π-orbitals provides stability of such formations. In the present study we investigated partially deuterated single crystal κ-(D$_4$-BEDT-TTF)$_2$Cu[N(CN)$_2$]Br, where the molecules forming dimers deployed with respect to one another at an angle of approximately 80°. In turn, dimers in the crystal form a layer in which they are packed in a triangular anisotropic lattice.

Organic salt κ-(BEDT-TTF)$_2$Cu[N(CN)$_2$]Br has an bilayer orthorhombic lattice (space group *Pnma*), wherein anions layers which are located in the *ac* plane are perpendicular to the *b*-axis (see Fig.1c), along which there are flat BEDT-TTF dimers (four per unit cell of the crystal lattice). The dimer planes are tilted along *a* - axis in opposite directions for adjacent layers. BEDT-TTF molecules have two stable conformers differing in orientation direction of the end of ethylene groups (eclipsed configuration if hydrogen atoms at the vertices of the hexagon are in the same directions, and staggered configuration in opposite case). At room temperature, these two conformations are experiencing continuous mutual transformation due to thermal fluctuations. By lowering the temperature the thermal motion of ethylene groups freezes, and BEDT-TTF molecules form a glassy phase. The speed of the cooling in the region of the glass transition ($T_g \approx$ 65-85 K) plays a decisive role in the residual disordering of molecules BEDT-TTF, which is stored then in the crystal to very low temperatures [5-6]. Unpaired π-electrons of skeleton of BEDT-TTF molecules provide the ability to charge transfer both within the dimer and between dimmers. A prerequisite for the charge transport in the crystal is the presence of anionic (Cu[N(CN)$_2$]Br) layers. Anion layer is composed of polymer zigzag chains, elongated in the direction of *a*, which consist of copper ions Cu+, linked by two dicyanamide bridges [(NC)N(CN)] - (DCA) and a halogen atom (Br) at the end (see Fig.1b). Anions, performing the function of an electron acceptor, can affect the packing of the donor molecule BEDT-TTF, which determines the nature of the transport properties, but do not directly participate in the conduction process. The electronic state of cation - anion system is interpreted as a simple effectively half-filled Mott-Hubbard state, and is determined by a subtle balance of band energy *W*, indoor/outdoor Coulomb energy of dimer (*U* and *V*, respectively), and the electron-phonon interaction between π-electrons existing in the low-dimensional lattice. If *U* is greater than *W*, the system goes into Mott's insulating state (AFM ground state). On the other hand, in the region where *W* exceeds *U*, the ground state is strongly correlated metallic state. A relatively high temperature superconductivity ($T_c \approx$ 10 K) is observed in this area [7-9]. A large impact on this ratio can have pressure applied to the crystal, which helps to reduce the ratio of *U/W*, and the type of halogen in the anion molecule, which increasing in mass has an effect comparable to the application of external pressure [10]. Small changes in pressure, temperature, chemical composition, have a decisive effect on the properties of such systems. For such objects with many degrees of freedom, the proper interpretation of the experimental results is often complex and ambiguous. The combination of structural and thermal methods of investigation, in particular, studies of thermal expansion substantially able to facilitate the interpretation of results, since various processes of magnetic and structural ordering appear differently in the physical properties of these substances. Precision measurement of thermal expansion and the specific heat at low temperatures can also provide information about the specific phase transitions induced by quantum phenomena at low energies. The aim of this work was to study the thermal expansion features of the single crystal κ-(D$_4$-BEDT-TTF)$_2$Cu[N(CN)$_2$]Br in a wide temperature range.

## 2. The thermal expansion of the sample κ-(D$_4$-BEDT-TTF)$_2$Cu[N(CN)$_2$]Br.

We curried out the thermal expansion measurement of monocrystalline sample κ-(D$_4$-BEDT-TTF)$_2$Cu[N(CN)$_2$] Br in the temperature range 2-290 K using a low-temperature high-sensitivity capacitive dilatometer [11]. The rate of sample cooling from room temperature to 2 K amounted to an average 0,74 K min$^{-1}$. Measurements of thermal expansion were carried out along the crystallographic direction *b* (Fig. 2), perpendicular to the crystal plane.

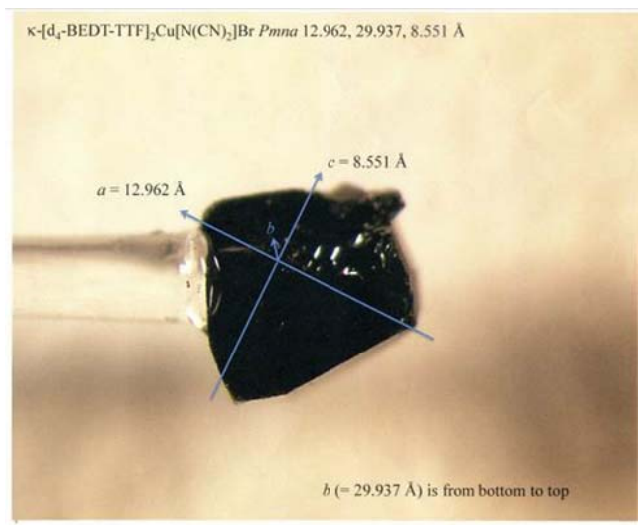

Fig. 2. Photopicture of the monocrystal used for this work. Lines show the crystallographic axes direction, obtained from the x-ray study.

The temperature dependence of the linear thermal expansion coefficient (LTEC) of the investigated sample is shown in the Fig. 3. It has a number of features. Below $T_c$ = 11,6 K [6] the LTEC of the sample has a small negative values (~ 10$^{-7}$ K$^{-1}$, Fig. 3c), which is apparently due to the transition from the paramagnetic metal in the superconducting state (see Fig. 4). Such behavior of LTEC when passing through $T_c$ had been observed previously in cuprates [13].

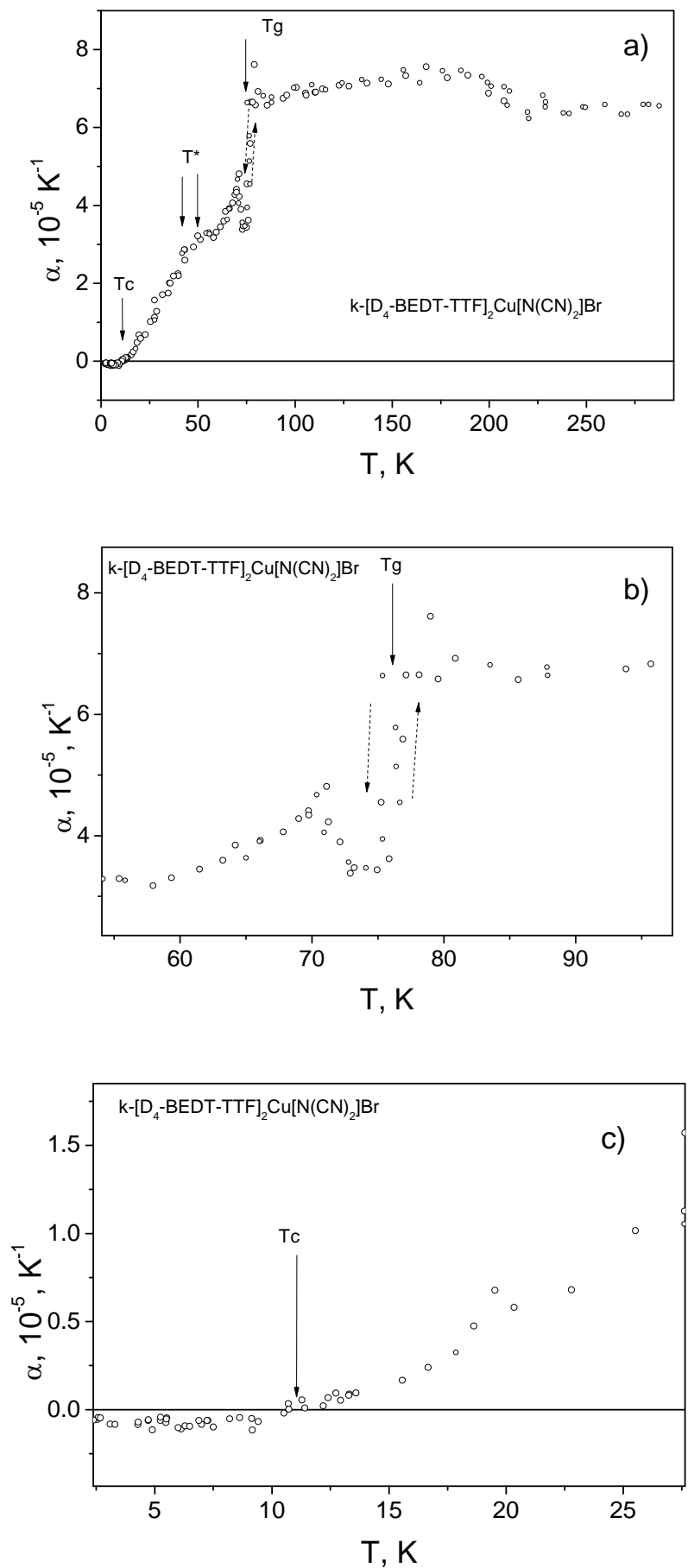

Fig. 3. The dependence of LTEC on temperature a) 2-290 K; b) 65-95 K; c) 2-30 K.

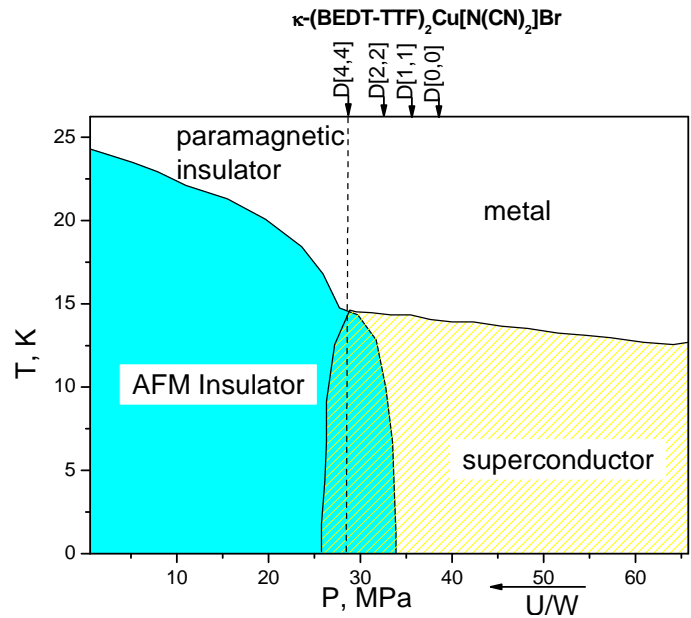

Fig. 4. Phase diagram of the dimer-Mott system based on κ-type salts consisting of BEDT-TTF molecules and effect of the deuteration degree (based on the data [12]).

In the temperature range of $T^* \sim 30$-$45$ K (Fig. 3a) there is a bend of the temperature dependence of the LTEC with a peak of 35-40 K. This is not considered as a phase transition but reflecting a gradual change in the lattice. In the vicinity of the temperature $T_g \sim 75$-$77$ K there is a thermal expansion anomaly (Fig. 3b), typical for the phase transition, a sharp jump in the LTEC values and hysteresis: the values obtained during the heating and cooling of the sample differed significantly. This anomaly is likely explained by the transition upon cooling of the sample in a glass-like state. A single point of view on the nature of this state is absent. Number of authors attributed this anomaly to the conformational freezing of the terminal ethylene moiety in BEDT-TTF molecules (see Fig.1d,f) [4-6]. An alternative view is that structural transformations of the polymeric chain will translate into changes of the orbital overlap determining the electronic properties within the BEDT-TTF layers [14].

At higher temperatures in the range 160-190 K, there is a maximum of LTEC (Fig. 3a), likely due to the influence of several competing mechanisms. Positive contribution to the thermal expansion makes the anharmonicity of the vibrations of BEDT-TTF dimers. The negative contribution may be related to the excitation of transverse optical vibrations of Cu-dicyanamide-Cu anionic polymer molecules chains. Reduction of the LTEC values with increasing temperature above 190 K, is apparently due to the increase of this contribution.

## 3. Influence of deuteration on the thermal expansion.

Investigated κ-(D$_4$-BEDT-TTF)$_2$Cu[N(CN)$_2$]Br sample was partially deuterated - four of the eight hydrogen atoms in ethylene BEDT-TTF groups have been replaced with deuterium. It is known that deuteration has a strong effect on the electronic state. κ-(D$_4$-BEDT-TTF)$_2$Cu[N(CN)$_2$]Br is located in the border region of the superconducting and antiferromagnetic phase, while D$_8$- compound and H$_8$- compound is in the antiferromagnetic and superconductiing phase, respectively [12].

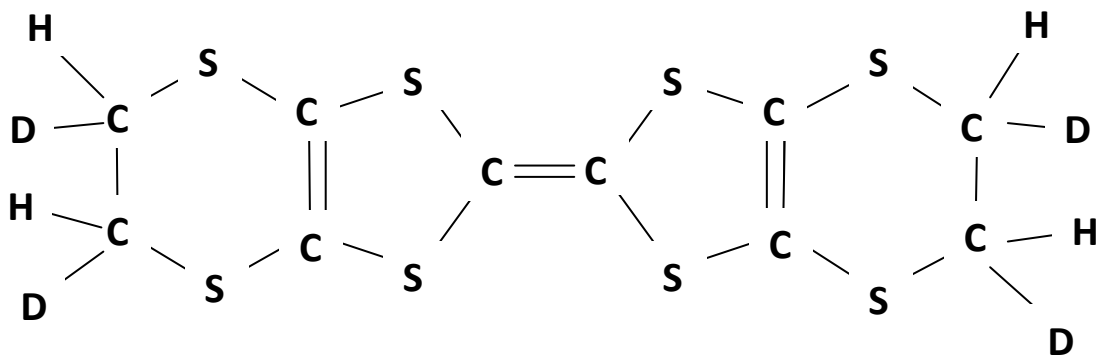

Fig. 5. Schematic image of the molecule d[2,2] BEDT-TTF [12].

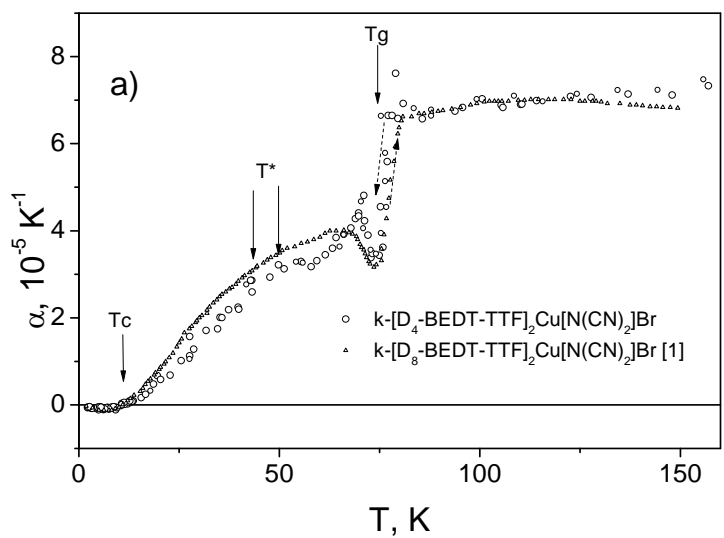

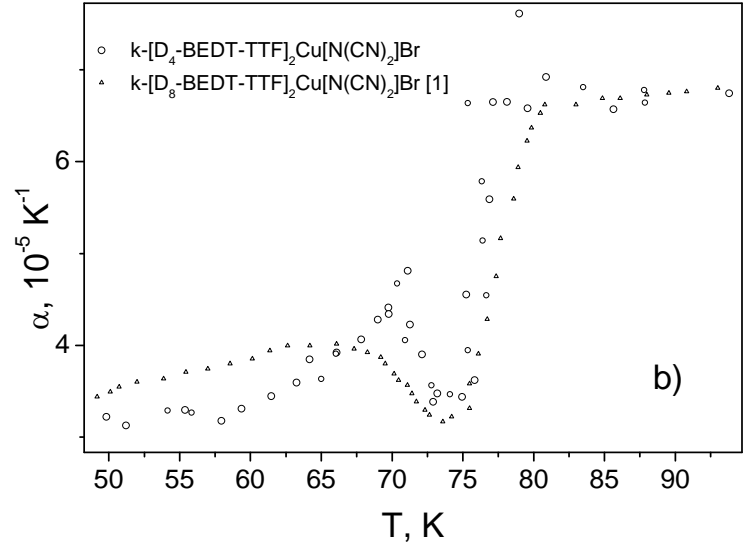

Fig. 6. Effect of the deuteration degree on thermal expansion of κ-(BEDT-TTF)$_2$Cu[N(CN)$_2$]Br.

Fig. 6 indicates that the deuteration degree of the sample (d [2;2] blank circle - this sample, d [4;4] triangles – data [6]) strongly affects the glass transition temperature $T_g$. For our sample d [2; 2] region of the glass phase transition was shifted toward lower temperatures (Fig. 6b), which is likely due to the smaller total angular momentum of the ethylene groups of molecules BEDT-TTF in the case of the sample d [2;2] compared with fully deuterated sample d [4;4], resulting in disinhibition of the synchronized motion of ethylene groups in both side of BEDT-TTF molecules. At the same time, changes in the deuteration degree weakly affected the transition temperature to the superconducting state $T_c$. It is further interesting that in the sample d [2;2] the anomaly at $T^*$ is much more pronounced and becomes sharper than for the sample d [4,4] (Fig. 6). This is considered as a change of elastic feature occurs in the Mott-Hubbard physics in organic systems. Kagawa et al observed by the transport measurements under precisely controlled helium gas pressures that the Mott boundary in the dimer-Mott system gives unusual criticality [15]. They further indicated that the end-point of the Mott boundary exits at about 33-38 K in the case of κ-(BEDT-TTF)$_2$Cu[N(CN)$_2$]Cl by the gas pressure controlled experiments to adjust $U/W$ ratio at the boundary. The higher temperature region than this critical end point temperature, a supercritical region like liquid-gas system under high pressure can exist. Since the d[2,2] sample exists just close to the boundary at ambient pressure, the observed pronounced anomaly may be related to the criticality around this end point. The elastic change occurs around the boundary region for κ-(BEDT-TTF)$_2$Cu[N(CN)$_2$]Cl compound under gas pressure is reported in [16]. More detail experiments around this region for d[2,2] sample and other deutrated samples with different ratio are necessary to investigate this point more clearly.

The formation of SDW like anomaly previously reported in this temperature region in κ-(BEDT-TTF)$_2$Cu[N(CN)$_2$]Cl may be related to this criticality and elastic feature. The observed hump like structure of LTEC and the difference of d[2,2] and d[4,4] are consistent with picture of the two-dimensional Mott boundary (fig.4).

In the temperature range 11-66 K the LTEC values for partially deuterated sample were systematically lower than for fully deuterated sample. In the case of deuterated sample, hydrogen bonding between donors and anions may be suppressed by the deuteration and the thermal expansion may become easier.

## 4. Conclusion.

Thermal expansion of monocrystalline sample κ-(D$_4$-BEDT-TTF)$_2$Cu[N(CN)$_2$] Br was studied along the crystallographic direction *b* (perpendicular to the plane of the crystal) in the temperature range 2-290 K. The temperature dependence of the linear thermal expansion coefficient (LTEC) of the investigated sample had a number of features. Below T$_c$ = 11,6 K the LTEC of the sample had a small negative values which is apparently due to the transition from the paramagnetic metal in the superconducting state. In the temperature range *T*\*~35-40 K there was the bend of the temperature dependence of the LTEC, which can be explained by the criticality of the Mott-Hubbard physics in 2D dimer-Mott state. A sharp jump in the LTEC values and hysteresis was observed in the vicinity of the temperature *T$_g$*~75-77 K, what is likely explained by the transition upon cooling of the sample in a glass-like state. A broad maximum of LTEC took a place in the range 160-190 K. Presumably, this feature is associated with the competing influence of the positive contribution to the thermal expansion caused by anharmonicity of vibrations of the BEDT-TTF dimers, and the negative contribution associated with the excitation of transverse optical vibrations of molecules of anionic polymer chains Cu-dicyanamide-Cu. Lowering the LTEC values with increasing temperature above 190 K is most likely due to increased role of the last contribution.

The thermal expansion of the sample κ-(D$_4$-BEDT-TTF)$_2$Cu[N(CN)$_2$]Br clearly manifested isotope effect in comparison with fully deuterated sample. In κ-(D$_4$-BEDT-TTF)$_2$Cu [N(CN)$_2$]Br the area of the glass phase transition was shifted toward lower temperatures. This can be due to smaller total rotational momentum of BEDT-TTF ethylene groups in the case of partially deuterated sample. In the temperature range 11-66 K values of LTEC for partially deuterated sample were systematically smaller than LTEC of fully deuterated sample. This can be attributed to the presence of a negative contribution due to quantum effects available in the energy spectrum of the partially deuterated molecule BEDT-TTF. For heavier fully deuterated BEDT-TTF, this effect is less pronounced.

**Acknowledgement**

The authors thank Prof. A. Kawamoto at Hokkaido University and Prof. H. Taniguchi at Saitama University for their synthesis of partially deuterated molecules. The authors are grateful Prof. A. Prokhvatilov for fruitful discussion of result.